\documentstyle[12pt,epsf]{article}
\newcommand{\beq}{\begin{equation}}
\newcommand{\eeq}{\end{equation}}
\newcommand{\beqa}{\begin{eqnarray}}
\newcommand{\eeqa}{\end{eqnarray}}
\newcommand{\im}{\mbox{Im}}
\newcommand{\re}{\mbox{Re}}

\begin{document}

\thispagestyle{empty}
\begin{flushright}
SLAC-PUB-7275\\
FERMILAB-PUB-96/308-T\\
hep-ph/9609357\\
September 1996
\end{flushright}
\vspace*{2cm}
\centerline{\Large\bf Mixing-induced CP Asymmetries}
\centerline{\Large\bf in Inclusive $B$ Decays
\footnote{
Research supported by the Department of Energy under contract
DE-AC03-76SF00515.}}
\vspace*{1.5cm}
\centerline{{\sc M. Beneke$^1$}, {\sc G. Buchalla$^2$}
and {\sc I. Dunietz$^2$}}
\bigskip
\centerline{\sl $^1$Stanford Linear Accelerator Center,}
\centerline{\sl Stanford University, Stanford, CA 94309, U.S.A.}
\vskip0.6truecm
\centerline{\sl $^2$Theoretical Physics Department}
\centerline{\sl Fermi National Accelerator Laboratory}
\centerline{\sl P.O. Box 500, Batavia, IL 60510, U.S.A.}

\vspace*{1.5cm}
\centerline{\bf Abstract}
\vspace*{0.2cm}
\noindent
We consider CP violating asymmetries that are induced by
particle-antiparticle mixing in inclusive 
channels of neutral $B$ meson decay. Not only are the branching
ratios sizable, at the 1\% to 50\% level, but some of those
asymmetries are expected to be large because of substantial CKM phases.
The inclusive sum partially dilutes the asymmetries, but the
dilution factor is calculable,
assuming local quark-hadron duality, and CKM parameters can be
reliably extracted. We discuss in detail the determination of 
$\sin 2\alpha$ from charmless final states in decays of $B_d$ mesons
and survey the asymmetries for other inclusive final states.
While probably not yet sensitive to standard model predictions, 
meaningful CP violation studies can be conducted with existing data
samples of inclusive neutral $B$ decays.
\vspace*{0.5cm}
\noindent

\vspace*{1.2cm}

\vfill

\newpage
\pagenumbering{arabic}

\section*{\normalsize\bf 1.~Introduction}
\label{intro}

Studying CP asymmetries in $B$ decays promises to reveal
crucial information about quark mixing and the nature of CP
violation. A rich phenomenology is expected and has been
widely discussed in the literature.
With the exception of a few `gold-plated' modes, it is
not yet clear which observables will eventually
turn out to be the optimal measures of quark mixing parameters.
For this reason, and in order to overconstrain the unitarity
triangle, it is important to consider different options.

In this letter we propose a class of CP violating asymmetries
that occur in partially inclusive neutral $B$ meson decays, that
is, in final states specified by their flavor content. Inclusive
asymmetries have been investigated earlier, for charged $B$ decays
\cite{charged}
which require a strong interaction phase difference, or for the dilepton
asymmetry in semileptonic decays of neutral $B$ mesons
\cite{dilepton}. Except for a few scattered studies
\cite{sachsd,aleksan,paschos}, mixing-induced asymmetries in 
inclusive nonleptonic neutral $B$ decays have not been analyzed so far, 
probably because they were thought to be very difficult to measure. 
In this note we wish to make the point that some inclusive asymmetries 
are complementary to those in exclusive $B$ decays, and sometimes even
theoretically advantageous.  The relevant inclusive branching
ratios range between 1\% to 50\% of all neutral $B$ decays.  They are
orders of magnitude larger than their exclusive counterparts.
Thus their experimental feasibility
should not be discarded but seriously studied.

Exclusive modes to measure CKM angles are usually favored
because of their unique experimental signatures. Theoretically, however,
hadronic uncertainties are difficult to quantify, so that clean extractions
of CKM angles often imply reliance on flavor symmetries and the
ability to measure several modes simultaneously. An example of this type
is the determination of $\sin 2\alpha$ from $B_d\to \pi^+\pi^-$ decays
\cite{GL}.

On the other hand, excellent vertex technology, $p/K/\pi$ separation,
a reliable heavy flavor decay model and hermiticity of the detector
with regard to charged tracks from $b$-decay distinguishes among the
various underlying quark transitions and different $b$-hadron species,
respectively.  Such distinctions will enable us to observe the 
CP asymmetries in inclusive data samples.  
Constructing the required detectors should be possible as indicated
by currently operating or planned devices.
The potential rewards are highly promising. Not only would
experiments be able to probe inclusive CP asymmetries 
that are expected to be sizable, but new determinations of $|V_{ub}/V_{cb}|$
may also become feasible.  Furthermore, one would be able to
flavor-tag almost all $b$-hadrons \cite{disting}, which is 
so crucial for mixing-induced CP violation studies.

As for the inclusive asymmetries,
in contrast to those in exclusive decays,
many hadronic uncertainties cancel in the sum over all
final states of a particular flavor content. This cancellation relies
on local parton-hadron duality. The quality of this assumption,
as well as the numerical values of remaining hadronic parameters
like decay constants, will be tightly constrained by other measurements,
so that the accuracy of the theoretical prediction for inclusive CP
asymmetries can be cross-checked.

After deriving basic formulae for asymmetries in Sect.~2,
we discuss in detail, in Sect.~3, the determination of
$\sin 2\alpha$
from the inclusive final state $\overline{u} u\overline{d} d$ (to be
precise, no charmed particles and net strangeness zero) and compare it
with its exclusive analogue $B_d\to \pi^+\pi^-$. Other
inclusive final states can be treated analogously and we
summarize the corresponding results in Sect.~4.  Sect.~5
concludes and presents an outlook.

\section*{\normalsize\bf 2.~Inclusive CP asymmetries}
\label{basicform}

The weak interactions mix neutral $B$ mesons with their
antiparticles. The time dependence of
a state $B(t)$ ($\overline B(t)$) that began as a flavor eigenstate
$B$ ($\overline B$) at $t=0$, can be written as
\begin{equation}\label{bt}
B(t)=g_+(t)\ B - \frac{q}{p}\ g_-(t)\ \overline B,
\end{equation}
\begin{equation}\label{bbt}
\overline B(t)=g_+(t)\ \overline B - \frac{p}{q}\ g_-(t)\ B.
\end{equation}
Here
\begin{equation}\label{qop}
\frac{q}{p}=\frac{M^*_{12}-i\Gamma^*_{12}/2}{(\Delta M- i
\Delta\Gamma/2)/2}=\frac{M^*_{12}}{|M_{12}|} \left(
1-\frac{1}{2} a\right)~, \qquad
a=\im\frac{\Gamma_{12}}{M_{12}},
\end{equation}
where $M_{12}$ ($\Gamma_{12}$) is the off-diagonal element in the
mass (decay width) matrix of the $B-\overline B$ system
($|B\rangle=|1\rangle$, $|\overline B\rangle=|2\rangle$).
$\Delta M=M_H-M_L$, $\Delta\Gamma=\Gamma_H-\Gamma_L$ are the
differences in mass and decay rate between the mass eigenstates
$B_{H,L}=p B\pm q\overline B$. Note that the sign convention for
$\Delta\Gamma$ is opposite to \cite{BBD1}. We use the CP phase
conventions $CP|B\rangle=-|\overline B\rangle$ and
$CP(\overline db)_{V-A}~[CP]^{-1}=-(\overline bd)_{V-A}$.
The second expression for $q/p$ in (\ref{qop}) is valid to
first order in the small quantity
$\Gamma_{12}/M_{12}={\cal O}(m^2_b/m^2_t)$.
The time dependent functions $g_\pm(t)$ are given by
\begin{equation}\label{gpt}
g_+(t)=e^{-iMt-\frac{1}{2}\Gamma t}\left[
\cosh\frac{\Delta\Gamma t}{4} \cos\frac{\Delta Mt}{2}
+i \sinh\frac{\Delta\Gamma t}{4} \sin\frac{\Delta Mt}{2}\right],
\end{equation}
\begin{equation}\label{gmt}
g_-(t)=e^{-iMt-\frac{1}{2}\Gamma t}\left[
\sinh\frac{\Delta\Gamma t}{4} \cos\frac{\Delta Mt}{2}
+i \cosh\frac{\Delta\Gamma t}{4} \sin\frac{\Delta Mt}{2}\right].
\end{equation}
with $M$ ($\Gamma$) the average mass (decay rate) of $B_H$
and $B_L$.

Given the final state $f$, the asymmetry
\begin{equation}\label{atdep}
{\cal A}(t)=
\frac{\Gamma(B(t)\to f)-\Gamma(\overline B(t)\to \overline{f})}{
\Gamma(B(t)\to f)+\Gamma(\overline B(t)\to \overline{f})}
\end{equation}
measures CP violation. The time-integrated asymmetry will be denoted
by ${\cal A}$. 
Usually $f$ represents an exclusive
CP eigenstate as in the familiar cases of
$B_d(\overline B_d)\to J/\psi K_S$ or $B_d(\overline B_d)\to\pi^+\pi^-$.
In the following we let $f$ be an inclusive final state,
for example all final states with no charm particles and no
net strangeness in the
decay of $B_d$ ($\overline B_d$). This channel is based on the quark level
transitions $\overline b(d)\to\overline uu\overline d(d)$ or
$b(\overline d)\to u\overline ud(\overline d)$. Further examples will
be described in Sect.~4. The expressions for the time-dependent
decay rates are
\begin{equation}\label{gbtf1}
\Gamma(B(t)\to f)=|g_+|^2\Gamma_{f,11}+\left|\frac{q}{p}\right|^2
|g_-|^2\Gamma_{f,22}-2{\re}\left(\frac{q}{p}\, g^*_+ g_-\Gamma_{f,12}
\right),
\end{equation}
\begin{equation}\label{gbtf2}
\Gamma(\overline B(t)\to f)=
\left|\frac{p}{q}\right|^2 |g_-|^2\Gamma_{f,11}+
|g_+|^2 \Gamma_{f,22}-2{\re}\left(\frac{p}{q}\, g^*_+ g_-\Gamma_{f,21}
\right),\end{equation}
with $\Gamma_{f,ij}=\sum_k\langle i|f_k\rangle\langle f_k|j\rangle$.
The sum runs over all final states $f_k$ that
contribute to the partially inclusive channel under
consideration. 
For any final state $f$, $\Gamma_{f,ji}=\Gamma^*_{f,ij}$.
If the summation were extended to include all
final states, $\Gamma_{f,ij}$ would coincide with $\Gamma_{ij}$,
the full decay constant matrix of the $B-\overline B$
system. 
Time-dependent studies determine 
$|q/p|^2\Gamma_{f,22}/\Gamma_{f,11}$ and   
\begin{equation}\label{xirdef}
\xi\equiv\frac{M^*_{12}}{|M_{12}|}\frac{\Gamma_{f,12}}{\Gamma_{f,11}}.
\end{equation}
The parameter $|q/p|^2 = 1 - a + {\cal O}(a^2)$ can
be obtained from the dilepton asymmetry (or, more generally,
from the asymmetry of tagged neutral $B$ mesons decaying to wrong
flavor-specific modes due to $B^0 - \overline B^0$ mixing).
Theory predicts $a$ to be $< 10^{-2}~~[< 10^{-3}]$ for the $B_d$ $[B_s]$
meson.  

In the remainder of this section we restrict ourselves to self-conjugate
final states common to
both $B$ and $\overline B$, $\overline{f}=f$ in (\ref{atdep}).
The quantity 
\begin{equation}\label{rdef}
r=\frac{\Gamma_{f,22}}{\Gamma_{f,11}}-1,
\end{equation}
parametrizes direct CP violation and should agree with the 
corresponding inclusive
asymmetry in charged $B$ decays up to corrections of ${\cal O}(1/m^3_b)$.
Using (\ref{gbtf1}), (\ref{gbtf2}) the time-dependent asymmetry
(\ref{atdep}) is then given by
\begin{equation}\label{atdex}
{\cal A}(t)=\frac{{\im}\,\xi \sin\Delta Mt-
a\left(\sin^2\frac{\Delta Mt}{2}+\sinh^2\frac{\Delta\Gamma t}{4}-
\frac{1}{2}\sinh\frac{\Delta\Gamma t}{2}\, {\re}\,\xi\right)-
\frac{r}{2} \cos\Delta Mt}{\left(1+\frac{r}{2}\right)
\left(1+2\sinh^2\frac{\Delta\Gamma t}{4}\right)
-\sinh\frac{\Delta\Gamma t}{2}\, {\re}\,\xi},
\end{equation}
which is valid to first order in $a$ and
neglecting terms of ${\cal O}(a\cdot r)$ and
${\cal O}(a\,({\rm Im}\xi)^2)$. The time-integrated asymmetry reads
\begin{equation}\label{atiex}
{\cal A}=\frac{\frac{x}{1+x^2}{\im}\,\xi-a\left(
\frac{x^2}{2(1+x^2)}+\frac{y^2}{2(4-y^2)}-\frac{y}{4-y^2}{\re}\,\xi
\right)-\frac{1}{2}\frac{r}{1+x^2}}{\left(1+\frac{r}{2}\right)
\left(1+\frac{y^2}{4-y^2}\right)-
\frac{2y}{4-y^2}{\re}\,\xi},
\end{equation}
where $x=\Delta M/\Gamma$ and $y=\Delta\Gamma/\Gamma$. Neglecting $y$ in
the previous two equations is an excellent approximation for $B_d$ mesons
where $|y| \;\raisebox{-.4ex}{\rlap{$\sim$}} \raisebox{.4ex}{$<$}\; 0.01$. 
Even for $B_s$ mesons, where
$|y|\sim 0.15$ is predicted, $|y\,\re\xi|/2 < 0.04$. Neglecting
$y$ in both cases leads to
\begin{equation}\label{atsimp}
{\cal A}(t)=\frac{2}{2+r}\left[{\im}\,\xi \sin\Delta Mt-
a\sin^2\frac{\Delta Mt}{2} -
\frac{r}{2} \cos\Delta Mt\right]\qquad t\ll \frac{1}{\Delta\Gamma},
\end{equation}
\begin{equation}\label{asimp}
{\cal A}=\frac{x}{1+x^2}\ \frac{2}{2+r}\left[{\im}\,\xi-\frac{x}{2} a -
\frac{1}{2 x} r \right] ,
\end{equation}
where (\ref{atsimp}) does not apply to 
$t\;\raisebox{-.4ex}{\rlap{$\sim$}} \raisebox{.4ex}{$>$}\;1/\Delta\Gamma$, 
which can be relevant for $B_s$ mesons.
The second and third terms in square brackets could be measured
separately, from the dilepton asymmetry and direct
CP violation in charged $B$ decays, respectively. The first term is
specific to the partially inclusive final state for neutral $B$ mesons 
and it is the one of interest here. For the charmless final
state, it is much larger than the other two, as will be seen below.
Let us also mention that for an exclusive decay $\xi=(M^*_{12}
\langle f|\overline B\rangle)/(|M_{12}|\langle f| B\rangle)$ and
$1+r=|\xi|^2$, and the well-known expressions for asymmetries in
exclusive decays are recovered from
(\ref{atsimp}), (\ref{asimp}).

Finally we note that due to $B-\overline{B}$ mixing a time
dependent CP asymmetry persists, even when all final states
are summed over. This asymmetry allows a direct determination of
$a$ since in that case $r=0$, ${\rm Im}\xi=x a/2$ and
${\rm Re}\xi=y/2$ in (\ref{atdex}), (\ref{atsimp}).

\section*{\normalsize\bf 3.~$\sin 2\alpha$ from charmless
inclusive $B_d$ decay}
\label{buud}

In this section we compute $\im\,\xi$ that enters the asymmetries
(\ref{atsimp}), (\ref{asimp}) for
the inclusive final state $f$ with no charmed particles
and net strangeness zero ($C=S=0$ with the additional constraint
of no $\overline{c} c$ pairs). We then discuss the determination
of $\sin 2\alpha$ from this channel. The quantity $\Gamma_{f,12}$ is
given by
\begin{equation}
\Gamma_{f,12}=\frac{1}{2M_B}\sum\!\!\!\!\!\!\!\int\limits_{k}\,\,\,
(2\pi)^4 \delta^{(4)}(p_B-p_{f_k})\,\langle B|{\cal H}_{eff}^\dagger|f_k
\rangle\langle f_k |{\cal H}_{eff}|\overline B\rangle.
\end{equation}
To lowest order in the strong interaction, the above
final state is uniquely associated with the $b\to u\overline{u} d$ transition
(and its hermitian conjugate) in the weak effective Hamiltonian
${\cal H}_{eff}$, up to penguin-penguin interference contributions,
which do not contribute to $\im\,\xi$, see below. Thus, $f$ is mainly
$\overline{u} u\overline{d} d$ (plus light quark pairs).
We may use completeness of the intermediate states to write 
(the approximate relation will be explained shortly)
\begin{equation}
\label{gfdef}
\Gamma_{f,12}\approx\frac{1}{2M_B}\langle B|\int d^4x\,
{\cal H}_{eff}^{f\dagger}(x){\cal H}_{eff}^f(0)|\overline B\rangle.
\end{equation}
The optical theorem further implies 
\begin{equation}
\label{tdef}
\Gamma_{f,12}\approx\frac{1}{2M_B}\langle B|{\im}\ i\, \int d^4x\,
T\,{\cal H}_{eff}^{f}(x){\cal H}_{eff}^f(0)|\overline B\rangle,
\end{equation}
where the relevant effective hamiltonian reads
\begin{equation}\label{huud}
{\cal H}_{eff}^f=\frac{G_F}{\sqrt{2}}\left[\lambda_u\left(
C_1 Q^u_1+C_2 Q^u_2\right)-\lambda_t\sum^6_{i=3}C_i Q_i\right]
+ \rm{h.c.}
\end{equation}
Here $\lambda_i=V^*_{id}V_{ib}$. The operators $Q^u_1$, $Q^u_2$ denote
`current-current' operators with flavor
content $(\overline{d} b)(\overline{u} u)$
and $Q_i$, $i=3,..,6$, denote `penguin' operators of the same flavor
content. The detailed expressions as well as the Wilson coefficient 
functions can be found in \cite{BBL}.

The right hand side of (\ref{tdef}) is related to the forward
scattering amplitude which can be expanded in the heavy quark mass,
following the methods reviewed in \cite{BIG}. Assuming only local
duality, this procedure allows us to go beyond the purely partonic
prediction, which is recovered as the leading term in the expansion.
On the other hand, the identification of $\Gamma_{f,12}$ with the
r.h.s. of (\ref{tdef}) is only approximate for the final state
with no charmed particles and therefore, strictly speaking, there is
no heavy quark expansion for the asymmetry. The identification would
be exact, if the final state were $C=S=0$, including $\overline{c} c$ pairs.
Higher order QCD effects mix the $C=S=0$ final states without and
with charmed particles (which, in our definition, include charmonia).
First, a gluon can be radiated in a $b\to u\overline{u} d$ transition and
split into a $\overline{c} c$. This is a very small correction, because it
requires a highly virtual gluon. Second, the $\overline{c} c$ pair,
created in the $b\to\overline{c} c d$ transition, can recombine and turn
into a $\overline{u} u$. While the leading logarithmic contribution from
this process is included in the $b\to d\overline{u} u$ penguin operators
above, the constant terms are not.  Since we only work to
leading logarithmic accuracy, it is consistent to neglect this
mixing. In the following we shall assume that both effects
are indeed small and treat (\ref{tdef}) as an equality.

Note that the problem just discussed does not exist for
$C=\pm 1$ final states
reached through the $b\to c$ or $b\to \overline{c}$ (plus light quarks
and perhaps a $\overline{c} c$ pair) transition, in which case
a heavy quark expansion is literally possible. We discuss this case
in Sect.~4.

After these general remarks, let us return to the calculation of
$\im\,\xi$. Combining (\ref{tdef}) and (\ref{huud}), we obtain
contributions from $Q_{1,2}$ interfering with themselves and from
penguin operators interfering with $Q_{1,2}$. The penguin-penguin
interference has CKM phase $\lambda^2_t/|\lambda_t|^2=\exp(2i\beta)$,
which is cancelled by the mixing phase $M^*_{12}/|M_{12}|=
\exp(-2i\beta)$. Therefore it does not contribute to ${\im}\,\xi$.
It is now straightforward to deduce $\Gamma_{f,12}$ from the
results of \cite{BBD1}, where $\Gamma_{12}$ in the $B_s-\overline B_s$
system has been computed\footnote{Eq.~(14) and Appendix A
of \cite{BBD1}. For $b\to u\bar ud$ and $B_d$ the CKM elements have 
to be adjusted in an obvious way and $z=m_c^2/m_b^2$ and $m_s$ are 
set to zero.}.

The total inclusive
decay rate of a $B_d$ meson into the $\overline uu\overline{d}d$ final state,
$\Gamma_{f,11}$ is given by
\begin{equation}\label{gf11}
\Gamma_{f,11}=\frac{G^2_F m^5_b}{192\pi^3}|V^*_{ud}V_{ub}|^2
\left(K_1+3 K_2\right),
\end{equation}
where
\begin{equation}\label{k1k2}
K_1=3 C^2_1+2C_1 C_2\approx -0.39 \qquad K_2=C^2_2\approx 1.25.
\end{equation}
Here we have neglected the small penguin contributions. We also omitted 
the known next-to-leading order radiative
corrections since these are not available for
$\Gamma_{f,12}$. For further use, we also define the combinations
\begin{equation}\label{k12pr}
K_1'=2(3 C_1 C_3+C_1 C_4+C_2 C_3)\approx 0.023\qquad
K_2'=2 C_2 C_4\approx -0.063,
\end{equation}
related to the interference of penguin operators with $Q_{1,2}$.
The numerical values of the coefficients $K_i$, $K_i'$ quoted
refer to evaluation of the Wilson coefficients at the scale
$m_b=4.8\,$GeV with $\Lambda_{QCD}=200\,$MeV. 
Next we recall that, within the phase conventions we are using,
the mixing phase for the $B_d$ system reads
$M^*_{12}/|M_{12}|=\exp(-2i\beta)$ and $\lambda_u/\lambda_u^*=
\exp(-2 i\gamma)$, where $\alpha,\beta,\gamma$ are the standard angles
of the unitarity triangle. Putting everything together we get
the result
\begin{eqnarray}\label{imxi}
{\im}\,\xi &=& -8\pi^2\,\frac{f^2_B M_B}{m^3_b}\,\sin 2\alpha\,
\Biggl[1+\frac{4}{3}\frac{2K_1+K_2}{K_1+3 K_2}\,(B-1)+
\frac{5}{3}\frac{K_2-K_1}{K_1+3 K_2}\,(B_S-1)
\nonumber\\
&&\hspace*{-1cm}-\frac{1}{3}\left(\frac{M^2_B}{m^2_b}-1\right)
+\,\frac{\sin\alpha\sin(\alpha+\beta)}{\sin\beta\sin 2\alpha}
\left(\frac{4}{3}\frac{2K'_1+K'_2}{K_1+3 K_2}\,B
+\frac{5}{3}\frac{K'_2-K'_1}{K_1+3 K_2}\,B_S\right)\Biggr].
\end{eqnarray}
Here $f_B$ is the $B_d$ meson decay constant in the normalization
in which $f_\pi=131\,$MeV and $M_B=5.28\,$GeV is the $B_d$
meson mass. The bag factors are defined in terms of hadronic matrix
elements as
\begin{eqnarray}\label{meqs}
\langle B|(\overline d_ib_i)_{V-A}(\overline d_jb_j)_{V-A}|\overline B\rangle
&=& \frac{8}{3} f^2_{B}M^2_{B}\,B,
\\
\langle B|(\overline d_ib_i)_{S+P}(\overline d_jb_j)_{S+P}|\overline B\rangle
&=& -\frac{5}{3} f^2_{B}M^2_{B}
\frac{M^2_{B}}{m^2_b}\,B_S.
\end{eqnarray}
$B=B_S=1$ corresponds to factorization (at the scale $m_b$). Notice
that in (\ref{imxi}), the scale-dependent coefficients $K_{1,2}$
enter only in the terms that deviate from the factorization limit.
This property, apparently an accident that would not persist beyond
the leading logarithmic approximation, also holds for the subleading
terms in the heavy quark expansion, proportional to $(M^2_B/m^2_b-1)
\sim \Lambda_{QCD}/m_b$. In evaluating
the hadronic matrix elements in these subleading terms we
have employed factorization at the scale $m_b$. More details
regarding their treatment can be found in \cite{BBD1}.
Retaining the complete ${\cal O}(\Lambda_{QCD}/m_b)$ corrections
has the advantage of avoiding various troublesome
ambiguities. For instance, working at leading order in the heavy
quark expansion it is not clear whether to use the bag parameters
and the decay constant in full QCD or those in the static limit,
which differ from the former by terms of
${\cal O}(\Lambda_{QCD}/m_b)$. Similarly
the numerically important distinction between the $b$-quark and
the $B_d$-meson mass cannot be made at leading order.
These problems are absent in (\ref{imxi}), where decay constant
and bag parameters have to be taken in full QCD, leading to an
expression that is complete to next-to-leading order in the heavy
quark expansion. Numerically, we find for $m_b=4.8\,$GeV
\begin{eqnarray}\label{num}
\im\,\xi &=& -0.12\,\sin 2\alpha\left(\frac{f_B}{180\,\mbox{MeV}}
\right)^2
\Biggl[1+0.19 (B-1)+0.81 (B_S-1)-0.07
\nonumber\\
&&\hspace*{-1cm}-0.05\,\frac{\sin\alpha\sin(\alpha+\beta)}
{\sin\beta\sin 2\alpha}\Biggr],
\end{eqnarray}
where we set $B=B_S=1$ in the penguin contribution.
With $x_d=0.73$, we see from (\ref{asimp}) that the
time-integrated asymmetry is
of order $5\%$ times $\sin 2\alpha$.

When the charm and up quarks
are degenerate in mass, all CP asymmetries must vanish in the CKM
model, while ${\cal A}$ in (\ref{asimp}) does not for the charmless
final state considered. There is no contradiction, because if
$m_c=m_u$, the asymmetry is no longer an observable, since the
charmless final state can not be experimentally distinguished from
a charmed final state. Summing also over final states with charm, the
asymmetry vanishes. Note that for charged $B$ decays,
even when $m_c\not= m_u$, the asymmetry for the inclusive $C=0$ final
state (and for the $C=\pm 1$ final states, see Sect. 4) 
must vanish by CPT conservation.
No such constraint exists for neutral $B$ decays due to 
$B-\overline{B}$ mixing.

Returning to (\ref{asimp}), we note that the terms involving the
direct CP asymmetry $r$ and the dilepton asymmetry $a$ do not
exceed several permille, as follows from the estimates of $r$ for
the $\overline{u} u \overline{d} d$, $\overline{d} d \overline{d} d$,
$\overline{s} s \overline{d} d$
final states in \cite{charged} and $a$ in \cite{dilepton}. Thus,
unless $\sin 2 \alpha$ is small, 
we may approximate ${\cal A}\approx \im\,\xi\cdot x_d/(1+x_d^2)$.
The penguin contribution to $\im\,\xi$ enters (\ref{imxi}), (\ref{num})
with small coefficient, but becomes enhanced if $\beta$ is close
to its current lower limit, $\beta\approx 0.18$. If in addition
$\alpha\approx \pi/2$, the penguin contribution can be sizable
and dominate the asymmetry. However, since the penguin term is 
calculable in the inclusive approach, it can be corrected for.
Note that the value of
$\beta$ that is needed for this purpose is related to the CP
asymmetry in the clean process $B_d(\overline B_d)\to J/\psi K_S$
and will be available when the inclusive asymmetry is measured.

The conventional method for the determination of $\sin 2\alpha$
makes use of the CP asymmetry in the exclusive decay
$B_d(\overline B_d)\to\pi^+\pi^-$. However, the
asymmetry coefficient (the factor multiplying the oscillation
term $\sin\Delta M t$) is not just $\sin 2\alpha$ but \cite{GRO1}
\begin{equation}\label{sin2p}
\sin 2\alpha- 2\left|\frac{A_2}{A_1}\right|\cos 2\alpha
\cos(\delta_1-\delta_2) \sin(\phi_1-\phi_2)
\end{equation}
where $A_i$, $\delta_i$ and $\phi_i$ are the amplitude, the strong
phase and the weak phase, respectively, of the tree ($i=1$) and
the penguin contribution ($i=2$) for $B_d\to\pi^+\pi^-$.
The strong phases and $|A_2/A_1|$ are unknown and
$\cos(\delta_1-\delta_2)$ could be one, in which case the
penguin contribution remains invisible in the time dependence.
In \cite{GRO2} it is argued that due to the penguin effects in
$B_d\to\pi^+\pi^-$ the asymmetry, which is supposed to measure
$\sin 2\alpha$, could be $0.4$ even if
$\sin 2\alpha\approx 0$.
The situation is less problematic for larger values of
$\sin 2\alpha$. It is possible to eliminate the
penguin contribution by an isospin analysis \cite{GL},
which requires the rates of $B^+\to\pi^+\pi^0$ and
$B_d\to\pi^0\pi^0$ and their CP conjugates as additional input.
The measurement of $B_d\to\pi^0\pi^0$ is experimentally extremely
challenging, in particular in view of the very small branching
fraction, which has been estimated to be below $10^{-6}$ \cite{KP}.

In contrast, the $B_d$ branching ratio governed by the
inclusive $b \to u \overline u d$ transitions is at the 1\% level,
orders of magnitude larger than the exclusive case.  The inclusive
CP asymmetry we are proposing thus offers an alternative
route towards measuring $\sin 2\alpha$. Unlike the exclusive case
it is not contaminated by the presence of unknown strong interaction
phases and the penguin contribution can be quantified. In addition,
it has the potential of becoming an
accurately known quantity as the knowledge of $f_B$, $B$ and $B_S$
improves, either through improvements in lattice gauge theory,
or additional measurements of mixing parameters. We also emphasize
that the assumption of local duality that underlies the
theoretical prediction can be checked by measuring the lifetime
difference of the $B_s$ mass eigenstates. If local duality
works for $\Gamma_{12}$ generated by the $b\to c\overline{c}s$ transition,
it will work only better for $b\to u\overline{u} d$. These advantages
are somewhat compensated by the dilution of the asymmetry incurred
by summing over many final states as well as the experimental challenge
of performing an inclusive measurement. But even if the situation turns
out to be more
favorable for $B\to\pi\pi$, the inclusive measurement would
provide a useful independent cross-check.

\section*{\normalsize\bf 4.~Inclusive CP asymmetries driven by other
quark transitions}
\label{other}

\begin{table}[t]
\addtolength{\arraycolsep}{0.0cm}
\renewcommand{\arraystretch}{1.2}
$$
\begin{array}{|c|c|c|c|c|}
\hline
\mbox{Final state} & \mbox{Transition} & d & \mbox{CKM factor} &
\mbox{Remarks} \\
\hline\hline
C=0,\,S=0 & \overline{b}\to\overline{u} u\overline{d} & & & \\[-0.2cm]
\mbox{no charm} & b\to u\overline{u} d
& \raisebox{1ex}[-1ex]{$-0.11$}
& \raisebox{1ex}[-1ex]{$\sin 2\alpha$}
& \raisebox{1ex}[-1ex]{$\mbox{(i)}$}\\ \hline
C=0,\,S=0& \overline{b}\to\overline{c} c\overline{d} & & & \\[-0.2cm]
\mbox{with charm} & b\to c\overline{c} d
& \raisebox{1ex}[-1ex]{$-0.41$}
& \raisebox{1ex}[-1ex]{$-\sin 2\beta$}
& \raisebox{1ex}[-1ex]{$\mbox{(ii)}$}\\ \hline
& \overline{b}\to\overline{c} u\overline{d} & & & \\[-0.2cm]
\raisebox{1ex}[-1ex]{$C=-1,\,S=0$} & b\to u\overline{c} d
& \raisebox{1ex}[-1ex]{$-0.21$}
& \raisebox{1ex}[-1ex]{$\lambda\,|V_{ub}/V_{cb}|\,\sin(\alpha-\beta)$}
& \raisebox{1ex}[-1ex]{$\mbox{(iii)}$}\\ \hline
& \overline{b}\to\overline{u} c\overline{d} & & & \\[-0.2cm]
\raisebox{1ex}[-1ex]{$C=1,\,S=0$} & b\to c\overline{u} d
& \raisebox{1ex}[-1ex]{$-0.21$}
& \raisebox{1ex}[-1ex]{$\lambda\,|V_{ub}/V_{cb}|\,\sin(\alpha-\beta)$}
& \raisebox{1ex}[-1ex]{$\mbox{(iii)}$}\\ \hline
\end{array}
$$
\caption{\label{table1}
${\rm Im}\xi\equiv d\cdot({\rm CKM\ factor})$ entering
CP asymmetries in inclusive $B_d$ decay. The dilution factor $d$ is
obtained with $m_b=4.8\,$GeV, $m_c=1.4\,$ GeV,
renormalization scale equal to $m_b$,
$f_B=180\,$MeV, $B$=$B_S$=1. Remarks: (i) Penguin contribution should be
taken into account, see Sect.~3. (ii) Penguin contribution enters with
CKM combination $\sin(\alpha+\beta)\sin\beta/\sin\alpha$ and remains
small. (iii) $d$ does not include
$1/m_b$ corrections.}
\end{table}

In this section we discuss the inclusive final states specified by
$S=0$ and (a) $C=0$ with charmed particles (as opposed to
Sect.~3), (b) $C=-1$ and (c)
$C=1$ in decays of both $B_d$ and $B_s$ mesons. The branching fraction
for the first channel is about 1\% $[20\%]$ for $B_d$ $[B_s]$, while 
(b) and (c) taken together comprise about 50\% $[3\%]$ of all $B_d$
$[B_s]$ decays. 
We emphasize that a significant fraction of time-evolved $B_d$'s
($\sim 10\%$) are seen in channel (c) due to 
$B_d-\overline{B}_d$ mixing. This channel could be overlooked
if one focussed on the tiny unmixed $B_d$ rate governed by
$\bar b\to\bar uc\bar d$.

We write
\begin{equation}
\im\,\xi = d\cdot\mbox{CKM factor},
\end{equation}
and list both factors in Tab.~\ref{table1} for $B_d$
and Tab.~\ref{table2} for $B_s$. The asymmetry is then
obtained from (\ref{atdex}) -- (\ref{asimp})
(in the case of CP self-conjugate final states, for the general
case see below). In order
to obtain the `dilution factor' $d$ for
other values of the bag factors $B$ and $B_S$ than those used in the
tables, the given values can be multiplied by $B_S$ for a rough
estimate. For a different choice of decay constant $f_B$, we recall
that $d$ depends quadratically on $f_B$. The definition
of the Wolfenstein parameters $\lambda,\rho,\eta$ can be found, e.g.,
in \cite{BBL}.

Let us turn to case (a). For $B_d$ decay, the asymmetry measures
$\sin 2\beta$. The dilution factor is larger than for the charmless
final state, mainly because the total width $\Gamma_{f,11}$ is
phase space suppressed for the $b\to c\overline{c} d$ transition. We
find a sizable $\im\,\xi \approx 0.41\sin 2\beta$, and with 
(\ref{asimp}) a significant asymmetry
${\cal A}\approx 0.20\sin 2\beta$. The penguin contribution is below
$5\%$ and has been neglected. 
The quantity $r$ is expected to be numerically small \cite{charged}
and vanishes identically in leading log approximation.
The same final state for $B_s$ decays normally involves an
$s\bar{s}$ pair and leads to a Cabibbo-suppressed interference term
$\im\,\xi\approx -\lambda^2\eta$.

\begin{table}[t]
\addtolength{\arraycolsep}{0.0cm}
\renewcommand{\arraystretch}{1.2}
$$
\begin{array}{|c|c|c|c|c|}
\hline
\mbox{Final state} & \mbox{Transition} & d & \mbox{CKM factor} &
\mbox{Remarks} \\
\hline\hline
C=0,\,S=0 & \overline{b}\to\overline{c} c\overline{s} & & &  \\[-0.2cm]
\mbox{with charm} & b\to c\overline{c} s
& \raisebox{1ex}[-1ex]{$-0.51$} & \raisebox{1ex}[-1ex]{$2\lambda^2\eta$}
& \raisebox{1ex}[-1ex]{$\mbox{(i)}$} \\ \hline
& \overline{b}\to \overline{c} u\overline{s} & & & \\[-0.2cm]
\raisebox{1ex}[-1ex]{$C=-1,\,S=0$}& b\to u\overline{c} s
& \raisebox{1ex}[-1ex]{$-0.28$} & \raisebox{1ex}[-1ex]
{$-\eta$}
& \raisebox{1ex}[-1ex]{$\mbox{(ii)}$} \\ \hline
& \overline{b}\to \overline{u} c \overline{s} & & & \\[-0.2cm]
\raisebox{1ex}[-1ex]{$C=1,\,S=0$}& b\to c\overline{u} s
& \raisebox{1ex}[-1ex]{$-0.28$} & \raisebox{1ex}[-1ex]
{$-\eta$} &
\raisebox{1ex}[-1ex]{$\mbox{(ii)}$} \\ \hline
\end{array}
$$
\caption{\label{table2}
${\rm Im}\xi\equiv d\cdot({\rm CKM\ factor})$ entering
CP asymmetries in inclusive $B_s$ decay. The dilution factor $d$ is
obtained with $m_b=4.8\,$GeV, $m_c=1.4\,$ GeV,
renormalization scale equal to $m_b$,
$f_{B_s}=210\,$MeV, $B$=$B_S$=1. Remarks: (i) The penguin contribution
has approximately the same CKM phase. (ii) $d$ does not include
$1/m_b$ corrections.}
\end{table}

Cases (b) and (c) require some generalization of Sect.~2, because
the final state is not self-conjugate, $f\not=\overline{f}$ in
(\ref{atdep}). To be definite consider case (b). The decay
$B_d(t)\to f$ is related to the 
$\overline{b}\to\overline{c} u \overline{d}$ and
$b\to u\overline{c} d$ transitions in the weak effective hamiltonian,
not including the hermitian conjugates. We call this piece
${\cal H}_{eff}^f$. The $2\times2$ mixing matrix that corresponds to
this decay is denoted by $\Gamma_{f,ij}$ and determined by the
matrix elements of ${\cal H}_{eff}^{f\dagger}(x){\cal H}_{eff}^f(0)$
according to (\ref{gfdef}). The decay $\overline{B}_d(t)\to\overline{f}$ 
is governed by the $b\to c\overline{u} d$ and
$\overline{b}\to\overline{u} c \overline{d}$ transitions
and the corresponding mixing matrix $\Gamma_{\overline{f},ij}$ is
related to ${\cal H}_{eff}^{f}(x){\cal H}_{eff}^{f\dagger}(0)$.
Since ${\cal H}_{eff}^f$ is not hermitian, $\Gamma_f$ and
$\Gamma_{\overline{f}}$ are not equal, but both matrices are
hermitian.

 The final states with $|C|=1,S=0$ are governed by single CKM
combinations. Consequently $\Gamma_{f,11}=\Gamma_{\overline{f},22}$,
$\Gamma_{f,22}=\Gamma_{\overline{f},11}$. Furthermore, 
the off-diagonal elements of $\Gamma_f$ and
$\Gamma_{\overline{f}}$ coincide within the leading logarithmic
approximation. Using these relations, we
derive the asymmetries
\begin{equation}
{\cal A}(t) = \frac{2\,\im\,((M_{12}^*\Gamma_{f,12})/|M_{12}|)\,\sin
\Delta M t - a \Gamma_{f,22} (1-\cos\Delta M t)}
{(1+\cos\Delta M t)\Gamma_{f,11}+(1-\cos\Delta M t)\Gamma_{f,22}}
\qquad t \ll \frac{1}{\Delta\Gamma},
\end{equation}
\begin{equation}\label{intasym}
{\cal A} = \frac{2 x\,\im\left(\frac{M_{12}^*}{|M_{12}|} \Gamma_{f,12}
\right) - x^2 a \Gamma_{f,22}}{(2+x^2) \Gamma_{f,11} + x^2\Gamma_{f,22}}.
\end{equation}
When one sums over cases (b) and (c), the final state $C=\pm 1$ is
self-conjugate and (\ref{atsimp}), (\ref{asimp}) apply. Although
no new information is obtained from this final state, it has the
experimental advantage that 
the charge of the single charm quark need not be determined.
Establishing that one, and only one, charmed
hadron exists in the final state is sufficient.

Now for the final state with $C=-1$, $\Gamma_{f,11}\gg\Gamma_{f,22}$,
because $B_d$ decays through $\overline{b}\to \overline{c} u\overline{d}$,
while
$\overline{B}_d$ decays through the CKM suppressed channel
$b\to u\overline{c} d$. For the final state with $C=1$, the situation
is just opposite. Thus, we get
\begin{eqnarray}
{\cal A}\! &=&\! \frac{2 x}{2+x^2}\, \im\,\xi
\qquad C=-1, \\
{\cal A}\! &=&\! \frac{2}{x}\,\left[\im\,\xi-\frac{x}{2} a\right]
\qquad C=1, \\
{\cal A}\! &=& \!\frac{x}{1+x^2}\,\left[2\im\,\xi-\frac{x}{2} a\right]
\qquad |C|=1. \label{asc1}
\end{eqnarray}
Here $\im\,\xi$ is defined as 
$\im\,\xi=\im(M^*_{12}\Gamma_{f,12}/(|M_{12}|\Gamma_{f,11}))$
where $f=u\bar cd\bar d$.
It is identical for all cases and can be read off from
Tab.~\ref{table1}. 
Note that since here the CP conjugate of $f$ is not yet included
in the definition of the final state, an explicit factor of
2 appears in front of $\im\,\xi$ in (\ref{asc1}) which is absent
in (\ref{asimp}).
With $|V_{ub}/V_{cb}|=0.08,\lambda=0.22$, we estimate
$\im\,\xi\approx 0.004\,\sin(\beta-\alpha)$. Assuming that the
term involving $a$ can still be neglected, we see that for
$x=x_d=0.73$ the asymmetry is about five times larger for the
final state with $C=1$ compared to $C=-1$, but still smaller than
about $1\%$. Since $\beta$ will be known from $B_d\to J/\psi K$, the
angle $\alpha$ or, equivalently $\gamma$ can be extracted.

We remark that the ratio $|V_{ub}/V_{cb}|$ that enters this asymmetry can
in principle be obtained from the same measurement of the
inclusive $b\to c \bar{u} q$ and $b\to u\bar{c} q$ ($q$=$d,s$) transitions,
because
\begin{equation}\label{vubvcb}
\frac{\Gamma_{f,22}}{\Gamma_{f,11}} = \frac{\Gamma(B^0_q \to
\bar{f})}
{\Gamma(B^0_q \to f)} = \left|\frac{V_{ub}V_{cq}}{V_{cb}V_{uq}}
\right|^2.
\end{equation}
While with neutral $B$ mesons the CKM extraction requires time-dependent studies, it can also be performed using a ratio of inclusive
time-integrated rates of $B^\pm$ decays.
Since in leading order the phase space functions coincide for the
$b\to c \bar{u} q$ and $b\to u\bar{c} q$ transitions,
the calculable corrections to the above ratio arise only at order
$\alpha_s$ and $1/m_b^3$. (This is in contrast to the determination
of $|V_{ub}/V_{cb}|$ from the ratio of inclusive semileptonic rates
$\Gamma(b \to u \ell \overline \nu) / \Gamma(b \to c \ell \overline \nu)$,
where mass effects do not cancel at tree level.)

Turning to $B_s$ mesons, we wish to consider several scenarios, 
because the $B_s - \overline B_s$ mixing parameter $x_s$ is very
large:
(1) The $(\Delta m)_{B_s} t$
- oscillations can be resolved;
(2) Although the $(\Delta m)_{B_s} t$ - oscillations cannot be resolved,
the two exponentials $\exp(-\Gamma_H t)$ and $\exp(-\Gamma_L t)$ can be
distinguished;
(3) Only time-integrated measurements (with or without a cutoff) can
be performed.
Scenario (1) would be ideal, and would allow us to determine the
relevant CKM parameters (phases and ratio of magnitudes) from 
flavor-tagged time-dependent studies.  The CKM model predicts those
time-dependent CP asymmetries to be of ${\cal O}(10\%)$
for inclusive transitions governed by $b \to c \overline u s/
u \overline c s$ (Tab.~\ref{table2}). 
For $B_s$, $\Gamma_{f,22}$ is not CKM suppressed compared to
$\Gamma_{f,11}$ and should therefore not be neglected. The quantity
$\im\,\xi$ shown in Tab.~\ref{table2} is defined by
$\im\,\xi=\im(M^*_{12}\Gamma_{f,12}/(|M_{12}|\Gamma_{f,11}))$
with $f=u\bar cs\bar s$. Detailed expressions for the asymmetries
can be derived from (\ref{gbtf1}), (\ref{gbtf2}).
If $(\Delta m)_{B_s} t$ - oscillations cannot be resolved, one could still
extract CKM information from time-dependences of untagged data samples
[scenario (2)]~\cite{bsbsbar}. If only time-integrated measurements
can be performed  the asymmetries become very small, inversely
proportional to $x_s$. In this case, the asymmetries could still
provide constraints on either $x_s$ or CKM parameters.

\section*{\normalsize\bf 5.~Conclusion and Outlook}
\label{concl}

One cannot overemphasize the importance of experimentally distinguishing
among the various inclusive $b$-quark transitions.  Once shown to be
experimentally feasible, then (a) almost all $b$-hadrons could be
flavor-tagged; (b) inclusive direct CP violating effects in charged $
B$ mesons and/or $b$-baryons could be probed; (c) new determinations of
$|V_{ub}/V_{cb}|$ may become possible; (d) mixing-induced inclusive
CP asymmetries, predicted to be sizable, could be searched for.

Both time-dependent and
time-integrated studies can discover inclusive, mixing-induced 
CP violation. Whenever possible, time-dependent measurements should 
be pursued. While the experimental hurdles can be daunting, the advantages
are obvious. Compared to exclusive transitions, inclusive modes
have huge branching fractions, ranging between 1\% to 50\%, with optimal
asymmetries between ${\cal O}(1\%)$ and ${\cal O}(20\%)$.
The sum of modes governed by a $b$-quark transition dilutes the CP
violating effects. However, this dilution factor is calculable,
up to $B$ meson matrix elements of local operators. These remaining 
hadronic parameters are already being calculated on the lattice
and will be available more accurately in the future, either from
improved lattice determinations or from other measurements of
quark mixing parameters.

In particular, the asymmetry in $B_d$ meson decay into
charmless final states with no net
strangeness provides a determination of $\sin 2\alpha$, that, compared
to the determination from exclusive modes, does not suffer from
uncontrolled penguin contributions, or the requirement to measure
modes with tiny branching fractions. The inclusive asymmetry
in $B_d$ decays driven by $b\to c\overline{c} d$ is large and 
leads to an independent constraint on $\sin 2\beta$.
The total inclusive time-dependent asymmetry, using all
neutral $B$ decays, is non-zero in general and measures $a$.

Single charmed final states, into which essentially half of all 
$B_d$ mesons decay, are predicted to show CP asymmetries of up to 1\%. 
CP violating effects of ${\cal O}(1\%)$ and ${\cal O}(10\%)$
are predicted for $B_s$ decays governed
by $b\to c\bar{c} s$ and $b \to u \overline c s/c \overline u s$,
respectively, where both channels are a measure of $\eta$.

Meaningful results could already be obtained from existing data
samples collected at LEP, SLC and Fermilab. 
CP violation is demonstrated if the number of flavor-tagged events 
differs from its CP-conjugated counterpart. 
(Here events denote specific neutral $B$ decay
topologies optimally weighted by additional information.) 
While flavor-tagging is automatically accomplished at polarized $Z$ factories  \cite{atwood}, it causes some statistical loss at unpolarized
$Z$ or $\Upsilon(4S)$ factories and at hadron accelerators.
We are eager to learn about the CP information already contained
within existing data.
In the future, improved technology and dedicated experiments
should then allow to probe in detail interesting aspects of flavor physics 
with inclusive $B$ decays.
 
\vspace*{0.7cm}

\noindent {\bf Acknowledgements.}
We are grateful to R.G. Sachs for introducing us to inclusive CP violation, to R. Aleksan for discussions concerning the experimental feasibility of distinguishing the various inclusive transitions, and to J. Incandela
and F. Snider for emphasizing the importance of vertexing.
We also thank J. Butler, R. Dubois, D. Jackson and J. Lewis for discussions.
Fermilab is operated by Universities Research Association, Inc.,
under contract DE-AC02-76CHO3000 with the United States Department
of Energy.


\vfill\eject


\begin{thebibliography}{99}
\bibitem{charged} M. Bander, D. Silverman and A. Soni,
Phys. Rev. Lett. {\bf 43} (1979) 242; J.-M. G\'{e}rard and
W.-S. Hou, Phys. Rev. {\bf D43} (1991) 2909; H. Simma, G. Eilam
and D. Wyler, Nucl. Phys. {\bf B352} (1991) 367;
R. Fleischer, Z. Phys. {\bf C58} (1993) 483
\bibitem{dilepton} A. Pais and S.B. Treiman, Phys. Rev. {\bf D12}
(1975) 2744; T. Altomari, L. Wolfenstein
and J.D. Bjorken, Phys. Rev. {\bf D 37} (1988) 1860;
M. Lusignoli, Z. Phys. {\bf C41} (1989) 645
\bibitem{sachsd}
I. Dunietz and R.G. Sachs, Phys. Rev.~{\bf D37} (1988) 3186;
[(E) ibid.~{\bf D39} (1989) 3515]
\bibitem{aleksan}
R. Aleksan et al., Phys. Lett. {\bf B317} (1993) 173;
Z.Phys. {\bf C67} (1995) 251; Phys. Lett. {\bf B356} (1995) 95
\bibitem{paschos}
A. Datta, E.A. Paschos and Y.L. Wu,
Nucl. Phys. {\bf B311} (1988) 35
\bibitem{GL}
M. Gronau and D. London, Phys. Rev. Lett. {\bf 65} (1990) 3381
\bibitem{disting}
I. Dunietz, FERMILAB-PUB-94/163-T, hep-ph/9409355
\bibitem{BBD1}
M. Beneke, G. Buchalla and I. Dunietz, FERMILAB-PUB-96/095-T,
hep-ph/9605259, {\it to appear in\/} Phys. Rev. {\bf D54}
\bibitem{BBL}
G. Buchalla, A.J. Buras and M.E. Lautenbacher,
"Weak Decays Beyond Leading Logarithms", FERMILAB-PUB-95/305-T,
hep-ph/9512380, {\it to appear in\/} Rev. Mod. Phys.
\bibitem{BIG}
I. Bigi {\it et al.},
{\it in} B Decays, second edition, ed. S. Stone
(World Scientific, Singapore, 1994)
\bibitem{GRO1}
M. Gronau, Phys. Rev. Lett. {\bf 63} (1989) 1451
\bibitem{GRO2}
M. Gronau, Phys. Lett. {\bf B300} (1993) 163
\bibitem{KP}
G. Kramer and W.F. Palmer, Phys. Rev. {\bf D52} (1995) 6411
\bibitem{bsbsbar}
I. Dunietz, Phys. Rev. {\bf D52} (1995) 3048
\bibitem{atwood}
W.B. Atwood, I. Dunietz and P. Grosse-Wiesmann, 
Phys. Lett. {\bf B216} (1989) 227;
W.B. Atwood, I. Dunietz, P. Grosse-Wiesmann, S. Matsuda and
A.I. Sanda, Phys. Lett. {\bf B232} (1989) 533

\end{thebibliography}
\end{document}